\documentclass{article}
\usepackage{spconf,amsmath,graphicx,hyperref}
\usepackage{amssymb}
\usepackage{multirow}
\usepackage{xcolor}

\title{Text-Prompted CLAP: Learning Text-Conditioned Audio Representations via Contrastive Learning}

\name{Mohan Li$^{\star \dagger}$ \qquad Rama Doddipatla$^{\dagger}$ \qquad Philip C. Woodland$^{\star}$}
\address{$^{\star}$ Department of Engineering, University of Cambridge, Cambridge, UK\\ $^{\dagger}$ Toshiba Cambridge Research Laboratory, Cambridge, UK}
\begin{document}
\ninept
\maketitle
\begin{abstract}
Contrastive Language–Audio Pretraining (CLAP) aligns text and audio in a shared embedding space, but encoding each modality independently limits its ability to model cross-modal semantics in complex audio understanding and retrieval tasks. To address this limitation, this paper proposes Text-Prompted CLAP (TP-CLAP), a parameter-efficient extension of CLAP that introduces a cross-attention-based fusion module to incorporate textual prompts into audio features. TP-CLAP is trained using an audio multiple-choice question answering (AMCQA) framework, where it learns to align text-conditioned audio representations with text embeddings of correct answer choices via contrastive learning. Experiments demonstrate that TP-CLAP matches or exceeds several substantially larger audio-LLMs on audio question answering despite its compact size. After fine-tuning for attribute-focused audio-to-audio retrieval, text-conditioned audio representations consistently outperform their unconditioned counterparts on music retrieval benchmarks. In addition, TP-CLAP improves upon the base CLAP model on conventional audio–text retrieval and zero-shot classification.

\end{abstract}
\section{Introduction}
\label{sec:intro}

Learning shared representations across modalities is a fundamental objective in multimodal machine learning. A major breakthrough in this area was Contrastive Language--Image Pretraining (CLIP) \cite{radford2021learning}, which demonstrated that large-scale contrastive learning \cite{chen2020simple} on paired image--text data can produce highly transferable representations and enable zero-shot generalisation across various downstream tasks. Inspired by this advance, Contrastive Language--Audio Pretraining (CLAP) \cite{elizalde2023clap} extended the same paradigm to the audio domain. By jointly training audio and text encoders to align semantically matched audio--text pairs in a shared embedding space, CLAP established strong performance for audio--text retrieval and zero-shot audio classification \cite{elizalde2023clap,elizalde2024natural,wu2023large, dinkel2026glap}.

Despite its success, the standard CLAP framework is limited in its ability to perform instruction-dependent tasks due to its independent encoding of audio and text. A notable example is audio question answering, where the model must selectively focus on aspects of the audio relevant to a given question, yet CLAP encodes audio without direct access to the question at all. A similar limitation arises in attribute-focused audio-to-audio retrieval, where the goal is to retrieve recordings that share a specific attribute (e.g., speaker emotion, acoustic environment, or instrumentation) with the query audio rather than based on global similarity. In both scenarios, the static audio representations produced by CLAP cannot readily adapt to emphasise such context-sensitive information.

Recent work has sought to address these challenges using pretrained multimodal architectures \cite{xiao2026scaling,chen2025wavrag,tang2025wave}. While these methods deliver promising results, they often rely on fine-tuning multimodal LLMs, leading to substantial computational cost during both training and inference. This raises an important question: \textit{Can text-conditioned audio representations be learned while preserving the efficiency and scalability of CLAP?}

To this end, this work proposes Text-Prompted CLAP (TP-CLAP)\footnote{\url{https://github.com/mohan-li/TP-CLAP}}, a lightweight extension of CLAP that incorporates textual prompts into audio representations through a cross-attention-based fusion module. The model is trained on audio multiple-choice question answering (AMCQA) data \cite{he2025measuring}, using questions as conditioning signals and aligning the fused embeddings with the text embeddings of the correct answer choices. This addresses audio question answering directly, and the same conditioning mechanism transfers to attribute-focused audio-to-audio retrieval through further fine-tuning on attribute-annotated data. The approach remains fully compatible with the original CLAP objective while introducing minimal architectural overhead, and it also improves performance on standard audio–text retrieval and zero-shot classification benchmarks.


\section{PROPOSED METHOD}
\label{sec:proposed}

\subsection{Standard CLAP Model}

CLAP jointly trains an audio encoder $E_a(\cdot)$ and a text encoder $E_t(\cdot)$ that map an audio clip $x_a$ and a text description $x_t$ into a common multimodal embedding space:
\begin{equation}
    \mathbf{e}_a = E_a(x_a), \quad \mathbf{e}_t = E_t(x_t).
\end{equation}
For notational convenience, a negative log-softmax over cosine similarities is
defined as
\begin{equation}
\ell(\mathbf{u}, \mathbf{v}^{+}, \mathcal{V}) = -\log
\frac{\exp(s(\mathbf{u}, \mathbf{v}^{+})/\tau)}
     {\sum_{\mathbf{v} \in \mathcal{V}} \exp(s(\mathbf{u}, \mathbf{v})/\tau)},
\label{eq:logsoftmax}
\end{equation}
where $s(\cdot)$ denotes cosine similarity and $\tau$ is a learnable
temperature that controls the sharpness of the similarity distribution.
Given a batch of $N$ paired audio--text samples, with
$\mathcal{E}_a = \{\mathbf{e}_a^j\}_{j=1}^{N}$ and
$\mathcal{E}_t = \{\mathbf{e}_t^j\}_{j=1}^{N}$ denoting the sets of audio
and text embeddings in the batch, CLAP is trained using a symmetric
InfoNCE loss to enforce bidirectional alignment:
\begin{equation}
\mathcal{L}_{\mathrm{CLAP}} = \frac{1}{2N} \sum_{i=1}^{N}
\Big[ \ell(\mathbf{e}_a^i, \mathbf{e}_t^i, \mathcal{E}_t)
    + \ell(\mathbf{e}_t^i, \mathbf{e}_a^i, \mathcal{E}_a) \Big].
\label{eq:clap}
\end{equation}

At inference time, cross-modal retrieval and zero-shot classification are performed by computing the cosine similarity between query and candidate embeddings. While this formulation is effective for learning general-purpose audio and text representations, the two modalities remain encoded independently throughout the process. Consequently, the resulting audio embeddings are text-agnostic and cannot adapt to textual context.

\subsection{Cross-attention Based Fusion Module}
\label{ssec:fusion_module}

\begin{figure}[t]
    \centering
    \includegraphics[width=0.75\columnwidth]{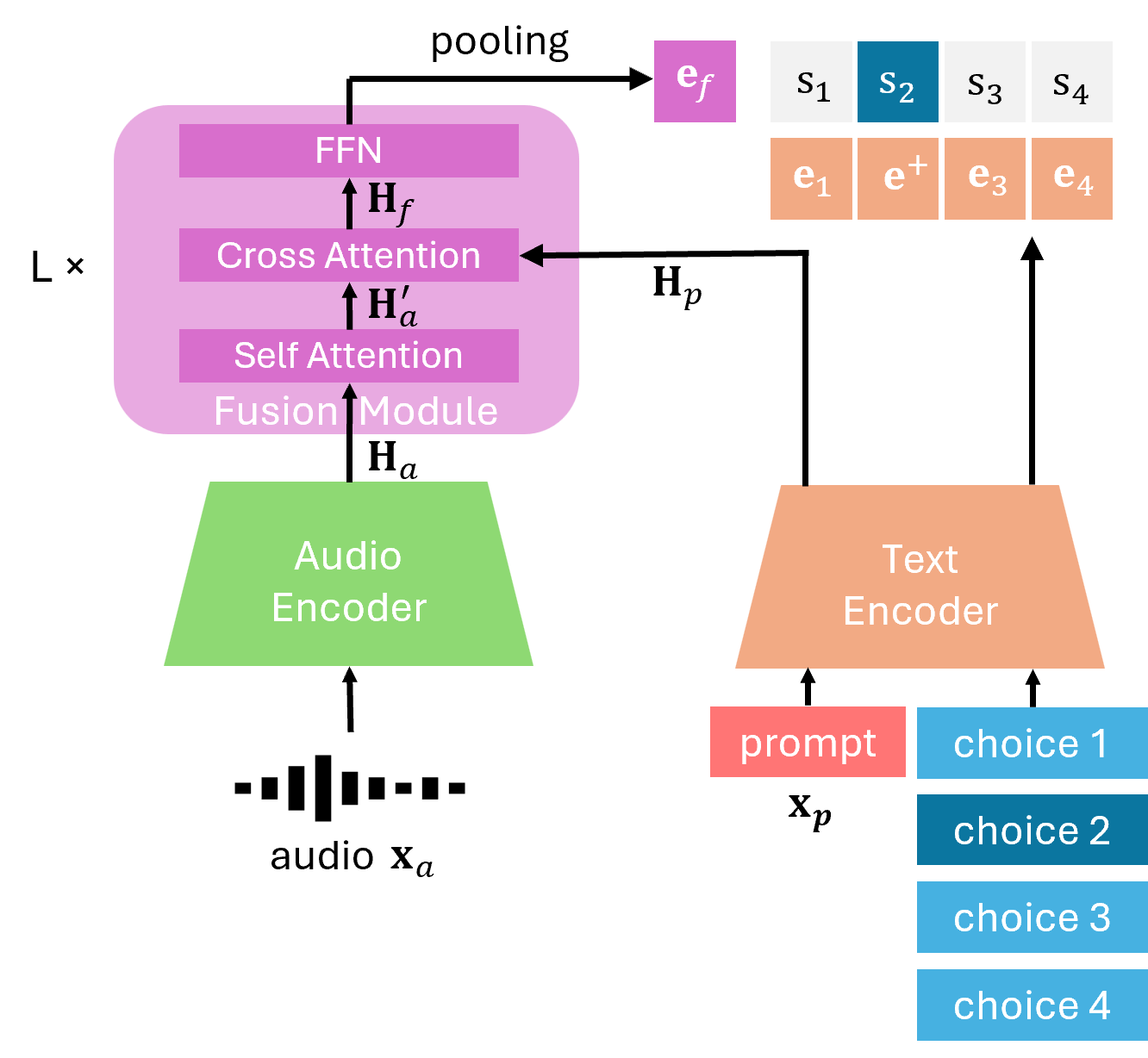}
    \vspace{-2mm}
    \caption{Architecture of the proposed TP-CLAP model.}
    \label{fig:tp-clap}
    \vspace{-2mm}
\end{figure}

To overcome this limitation, a fusion module $F(\cdot)$ is introduced which conditions audio representations on textual prompts (See Figure~\ref{fig:tp-clap}). Specifically, it builds upon a pretrained CLAP model and places a stack of cross-attention blocks on the audio encoder. Given an audio clip $x_a$ and a textual prompt $x_p$, the pretrained audio and text encoders first produce token-level hidden representations
\begin{equation}
\mathbf{H}_a \in \mathbb{R}^{T_a \times d}, \qquad
\mathbf{H}_p \in \mathbb{R}^{T_p \times d},
\end{equation}
where $T_a$ and $T_p$ are the numbers of audio and text tokens, respectively, and $d$ is the hidden dimension.

Each fusion block consists of a self-attention layer, a cross-attention layer, and a feed-forward network. Letting $\mathbf{H}^{'}_a$ denote the audio token representations after the self-attention layer, the cross-attention layer uses $\mathbf{H}^{'}_a$ as queries and the text representations $\mathbf{H}_p$ as keys and values:
\begin{equation}
\mathbf{H}_f =
\mathrm{CrossAttn}
(\mathbf{Q}=\mathbf{H}^{'}_a,
\mathbf{K}=\mathbf{H}_p,
\mathbf{V}=\mathbf{H}_p),
\end{equation}
where $\mathbf{H}_f$ denotes the resulting fused text--audio token representations. After the subsequent FFN, the final token representations are aggregated using a pooling layer to produce a fixed-dimensional embedding $\mathbf{e}_f$, which serves as the text-conditioned audio representation for downstream training and inference.

\subsection{AMCQA Supervision Framework}
\label{subsec:audio-mcq}

Audio multiple-choice question answering (AMCQA) data is leveraged as a natural extension of the CLAP training paradigm. Given an audio clip $x_a$, a question prompt $x_p$, and $M$ candidate answer choices $\{c_1,\ldots,c_M\}$, the fused embedding is produced as discussed in Section~\ref{ssec:fusion_module}:
\begin{equation}
\mathbf{e}_f = F(\mathbf{H}_a, \mathbf{H}_p).
\end{equation}
The text embeddings of the candidate answers are produced as
\begin{equation}
\mathbf{e}_m = E_t(c_m), \quad m = 1,\ldots,M,
\end{equation}
where $\mathbf{e}^+$ denotes the embedding of the correct answer choice.

An InfoNCE loss for the fused embedding is adopted which aligns $\mathbf{e}_f$ with $\mathbf{e}^+$ while contrasting it against distracting answer choices from both the current and other questions in the mini-batch:
%
\begin{equation}
\mathcal{L}_{\mathrm{InfoNCE}}^{f} = \frac{1}{N} \sum_{i=1}^{N}
\ell(\mathbf{e}_f^i, \mathbf{e}^{+,i}, \{\mathbf{e}_m^j\}_{j=1,\,m=1}^{N,\,M}),
\label{eq:infonce-f}
\end{equation}
where $\mathbf{e}^{+,i}$ is the correct answer embedding for the $i$-th question.

To further improve discrimination among candidates within each question, a cross-entropy loss is introduced:
\begin{equation}
\mathcal{L}_{\mathrm{CE}}^{f} = \frac{1}{N} \sum_{i=1}^{N}
\ell\big(\mathbf{e}_f^i, \mathbf{e}^{+,i},
\{\mathbf{e}_m^i\}_{m=1}^{M}\big).
\label{eq:ce-f}
\end{equation}
Now, the overall AMCQA objective for the fused embedding is defined as
\begin{equation}
\mathcal{L}_{\mathrm{AMCQA}}^{f}
=
\mathcal{L}_{\mathrm{InfoNCE}}^{f}
+
\lambda \mathcal{L}_{\mathrm{CE}}^{f},
\label{eq:loss_mcq}
\end{equation}
where $\lambda$ is the weight of the cross-entropy loss.

To maintain the general-purpose audio--text alignment learned during CLAP pretraining, two complementary regularisation terms are introduced. First, the standard CLAP loss $\mathcal{L}_{\mathrm{CLAP}}$ (Equation~\ref{eq:clap}) is retained by jointly training on standard audio--caption data. Second, the same AMCQA objective is applied to the unfused audio embedding $\mathbf{e}_a$, where text embeddings are computed from concatenated question--choice pairs rather than choices alone:
\begin{equation}
\tilde{\mathbf{e}}_m = E_t([x_p; c_m]).
\end{equation}
This yields an auxiliary loss $\mathcal{L}_{\mathrm{AMCQA}}^{u}$ for the unfused embedding, defined analogously to Equation~\ref{eq:loss_mcq}, but operating on $(\mathbf{e}_a, \tilde{\mathbf{e}}_m)$ pairs. As a result, the final training objective is
\vspace{-0mm}
\begin{equation}
\mathcal{L}
=
\mathcal{L}_{\mathrm{AMCQA}}^{f}
+
\alpha \mathcal{L}_{\mathrm{CLAP}}
+
\beta \mathcal{L}_{\mathrm{AMCQA}}^{u},
\label{eq:loss}
\end{equation}
where $\alpha$ and $\beta$ control the strengths of the regularisation terms.

\subsection{Attribute-Focused Audio-to-Audio Retrieval}

The AMCQA training stage equips TP-CLAP with the ability to inject textual prompts into audio representations. To adapt the model for attribute-focused audio-to-audio retrieval, it is further fine-tuned using audio classification data annotated with multiple attribute dimensions, such as genre, mood, or instrumentation. For each attribute, a predefined question (e.g., \texttt{``What is the genre of the music?''}) is used as the text prompt, enabling the model to produce attribute-specific audio representations.

To encourage the learned embedding to capture the target attribute, a classification objective is first applied. Let $\{l_1,\ldots,l_M\}$ denote the candidate labels for the attribute, with corresponding text embeddings
\begin{equation}
\mathbf{e}_m = E_t(l_m).
\end{equation}
The classification loss $\mathcal{L}_{\mathrm{cls}}$ is defined similarly to Equation \ref{eq:ce-f}.

In addition, an audio-to-audio contrastive objective is introduced. Rather than contrasting two text-conditioned embeddings, the loss is computed between the text-conditioned embedding $\mathbf{e}_f$ and the unconditioned embedding $\mathbf{e}_a$. This design allows each audio clip to be represented by a single embedding during retrieval, avoiding the need to pre-compute and store separate embeddings for different attributes in the database.

For a mini-batch of $N$ audio clips, an attribute similarity target $t_{ij}\in[0,1]$ is defined between samples $i$ and $j$. For single-label attributes,
\begin{equation}
t_{ij}
=
\begin{cases}
1, & \text{if the labels match},\\
0, & \text{otherwise}.
\end{cases}
\end{equation}
As for multi-label attributes, the target is computed using the intersection-over-union (IoU) between the corresponding label sets:
\begin{equation}
t_{ij}
=
\frac{|Y_i \cap Y_j|}
{|Y_i \cup Y_j|}.
\end{equation}
Then the audio-to-audio contrastive loss is given by
\begin{equation}
\mathcal{L}_{\mathrm{a2a}} = \frac{1}{N} \sum_{i=1}^{N} \sum_{j=1}^{N}
t_{ij}\, \ell(\mathbf{e}_f^i, \mathbf{e}_a^j, \mathcal{E}_a).
\label{eq:a2a}
\end{equation}
The final fine-tuning objective is
\begin{equation}
\mathcal{L}
=
\mathcal{L}_{\mathrm{cls}}
+
\gamma \mathcal{L}_{\mathrm{a2a}},
\end{equation}
where $\gamma$ controls the contribution of the contrastive objective.

\section{EXPERIMENTS}
\label{sec:exp}

\subsection{Implementation and Training Details}

TP-CLAP is built upon a CLAP backbone with \texttt{CED-Base} \cite{dinkel2024ced} and \texttt{bert-base-uncased} \cite{devlin2019bert} as the audio and text encoders, respectively, each followed by a 1024-dimensional projection head. The proposed fusion module consists of 2 cross-attention blocks (hidden size 768, 12 attention heads) followed by another 1024-dimensional projection head.

The base CLAP model is trained on AudioCaps-v2 \cite{kim2019audiocaps}, Clotho \cite{drossos2020clotho}, WavCaps \cite{mei2024wavcaps}, MACS \cite{martin2021ground}, LP-MusicCaps MC \cite{doh2023lp}, and LP-MusicCaps MTT \cite{doh2023lp}, totalling 601k audio--caption pairs. Training is performed for 20 epochs using the AdamW optimiser with a learning rate of $5\times10^{-5}$ and a 2-epoch warmup.

TP-CLAP is further trained on the sound and music subsets of AudioMCQ \cite{he2025measuring} (310k question--choice pairs) in two stages: (1) optimising only the fusion module for 5 epochs with a learning rate $5\times10^{-5}$ and a 1-epoch warmup while freezing the CLAP encoders; and (2) jointly fine-tuning all parameters for 1 epoch on both audio--caption and AudioMCQ data using a learning rate of $5\times10^{-6}$. A batch size of 512 is used throughout, and audio longer than 10 seconds is randomly cropped during training. Hyperparameters $\tau$ (initial), $\lambda$, $\alpha$, $\beta$, and $\gamma$ are set to 0.07, 0.5, 1, 1, and 1, respectively.

To isolate the effect of the fusion module, a baseline (CLAP+\allowbreak AudioMCQ) is constructed which retains the original CLAP architecture but is trained with the same data and hyperparameters as TP-CLAP using the objective of $\mathcal{L}_{\mathrm{CLAP}}+\mathcal{L}_{\mathrm{AMCQA}}^{u}$ (Section~\ref{sec:proposed}).

For attribute-focused audio-to-audio retrieval, TP-CLAP is fine-tuned on two music retrieval benchmarks: (1) NSynth \cite{engel2017neural}, using the instrument family, pitch, and instrument source attributes, comprising 870k audio--attribute pairs; and (2) MagnaTagATune \cite{law2007tagatune}, using the genre, instrument, and tempo attributes, comprising 24k pairs. The models are trained for 30 epochs with a learning rate of $5\times10^{-5}$ and batch sizes of 512 and 256, respectively.

\subsection{Evaluation}

The models are first evaluated on standard audio–text retrieval tasks using the AudioCaps and Clotho test sets. Following prior work, retrieval performance is reported using Recall@1 (R@1).

Zero-shot audio classification is then evaluated on eight benchmarks spanning sound (ESC50 \cite{piczak2015esc}, FSD50K \cite{fonseca2021fsd50k}, UrbanSound8K (US8K) \cite{salamon2014dataset}, VocalSound (VS) \cite{gong2022vocalsound}, and CREMA-D (CD) \cite{cao2014crema}) and music (GTZAN \cite{tzanetakis2002musical}, Beijing Opera (BO) \cite{tian2014study}, and NSynth instrument) domains. Mean average precision (mAP) is reported for FSD50K, and accuracy is used for all other datasets. Note that, for TP-CLAP, retrieval and classification use the underlying CLAP encoders rather than the fusion module.

Next, audio understanding is evaluated on the sound and music subsets of MMAU \cite{sakshi2025mmau} and MMAR \cite{ma2026mmar}. For vanilla CLAP and CLAP+\allowbreak AudioMCQ, answer scores are computed as the similarity between the unconditioned audio embedding $\mathbf{e}_a$ and the concatenated question--choice embedding $\tilde{\mathbf{e}}_m$. For TP-CLAP, this score is interpolated with the similarity between the text-conditioned audio embedding $\mathbf{e}_f$ and the choice embedding $\mathbf{e}_m$, using equal weights as in training. Results are reported in terms of accuracy.

Finally, attribute-focused audio-to-audio retrieval is evaluated on the test sets of NSynth and MagnaTagATune. For NSynth, retrieval quality is measured using Precision@$K$ (P@$K$), defined as the proportion of retrieved items sharing the target attribute with the query, and mAP. For MagnaTagATune, where attributes are multi-label, SoftPrecision@$K$ (SP@$K$) and SoftmAP (SmAP) are used to account for partial label overlap between retrieved items and queries.

\section{RESULTS}
\label{sec:results}

\subsection{Audio--Text Retrieval}


\begin{table}[t]
\centering
\caption{Audio--text retrieval results on AudioCaps and Clotho.}
\label{tab:retrieval_results}
\begin{tabular}{l|cc|cc}
\hline
\multirow{2}{*}{Model} &\multicolumn{2}{c|}{AudioCaps} & \multicolumn{2}{c}{Clotho} \\
& T2A & A2T & T2A & A2T \\ \hline
MS-CLAP \cite{elizalde2024natural} & 35.6 & 42.5 & 15.7 & 22.9  \\
Laion-CLAP \cite{wu2023large} & 35.1 & 44.2 & 16.9 & 24.4 \\
GLAP \cite{dinkel2026glap} & 41.7 & 54.4 & 19.4 & 21.8 \\ \hline
CLAP (ours) & 42.1 & 55.9 & 21.0 & 26.6 \\
CLAP+AudioMCQ & \textbf{42.7} & 56.1 & 21.2 & 26.8 \\
TP-CLAP (proposed) & 42.6 & \textbf{57.2} & \textbf{21.3} & \textbf{27.1} \\ \hline
\end{tabular}
\vspace{-3mm}
\end{table}

\begin{table*}[t]
\centering
\caption{Zero-shot audio classification results (\%). Numbers in grey are not obtained under a strict zero-shot setting.}
\label{tab:zs_results}
\begin{tabular}{l|ccccc|ccc|c}
\hline
\multirow{2}{*}{Model} &\multicolumn{5}{c|}{Sound} & \multicolumn{3}{c|}{Music} & \multirow{2}{*}{Avg.} \\
& ESC50 & FSD50K & US8K & VS & CD & GTZAN & BO & NSynth & \\ \hline
MS-CLAP \cite{elizalde2024natural} & \textbf{93.9} & \textcolor{gray}{48.5} & \textbf{82.3} & 80.0 & 30.0 & 58.4 & 46.6 & \textcolor{gray}{47.9} & \textcolor{gray}{61.0} \\
Laion-CLAP \cite{wu2023large,dinkel2026glap} & 91.0 & 21.5 & 77.0 & 79.3 & 18.3 & 47.4 & 40.2 & 26.1 & 50.1 \\
GLAP \cite{dinkel2026glap} & 88.8 & 40.9 & 78.9 & 75.1 & 20.5 & \textbf{69.6} & 36.5 & 31.3 & 55.2 \\ \hline
CLAP (ours) & 93.0 & 51.2 & 81.0 & 83.8 & 30.0 & 57.8 & 66.9 & 39.9 & 63.0 \\
CLAP+AudioMCQ & 93.0 & 52.1 & 81.9 & \textbf{84.1} & 33.1 & 56.9 & 65.3 & 40.6 & 63.4 \\
TP-CLAP (proposed) & 93.6 & \textbf{52.6} & 82.2 & \textbf{84.1} & \textbf{34.8} & 55.9 & \textbf{67.8} & \textbf{42.1} & \textbf{64.1} \\ \hline
\end{tabular}
\vspace{-6mm}
\end{table*}

Table~\ref{tab:retrieval_results} presents audio--text retrieval results on AudioCaps and Clotho. The proposed model is compared against representative CLAP-style baselines, including MS-CLAP~\cite{elizalde2023clap}, Laion-CLAP~\cite{wu2023large}, and GLAP~\cite{dinkel2026glap}, which are trained on substantially larger and more diverse audio--caption corpora than ours. Despite using much less pretraining data, our CLAP model outperforms these baselines by leveraging the powerful CED audio encoder. Building on this, training with additional AudioMCQ data yields consistent improvements over the base CLAP model, suggesting that the fine-grained semantic supervision provided by the task benefits general audio--language alignment. The proposed TP-CLAP achieves the best overall performance, obtaining the highest Audio-to-Text R@1 score on AudioCaps and the best results for both metrics on Clotho. The improvements over CLAP+AudioMCQ indicate that the gains cannot be attributed solely to the inclusion of AudioMCQ data, but also to the proposed cross-attention-based fusion mechanism, which enables more effective token-level interaction between audio and textual information during training.

\subsection{Zero-Shot Audio Classification}

Table~\ref{tab:zs_results} reports zero-shot classification results on eight sound and music benchmarks. Our CLAP model establishes a stronger baseline than the selected CLAP variants reported in the literature. Compared to the base CLAP model, adding AudioMCQ data improves the average score from 63.0\% to 63.4\%, suggesting that the question-answering task enhances the discriminative power of the learned audio--language representations. TP-CLAP further increases the average score to 64.1\%, achieving the best performance on FSD50K, VocalSound, CREMA-D, Beijing Opera, and NSynth instrument. These improvements are consistent with the retrieval results in Table~\ref{tab:retrieval_results}, indicating that both AMCQA supervision and the proposed fusion mechanism contribute to enhancing the transferability of the underlying CLAP representations.

\subsection{Audio Question Answering}

\begin{table}[t]
\centering
\caption{Audio question answering accuracy (\%) of audio-LLMs (top) and CLAP-based models (bottom) on MMAU and MMAR.}
\label{tab:audiomcq_sources}
\resizebox{\columnwidth}{!}{
\begin{tabular}{l|c|cc|cc}
\hline
\multirow{2}{*}{Model} & \multirow{2}{*}{Size} & \multicolumn{2}{c|}{MMAU} & \multicolumn{2}{c}{MMAR} \\
& & Sound & Music & Sound & Music \\ \hline
Audio Flamingo 3 \cite{ghosh2026audio} & 8B & \textbf{79.58} & \textbf{73.95} & \textbf{53.33} & \textbf{50.97} \\
Audio Flamingo 2 \cite{ghosh2025audio} & 3B & 71.47 & 70.96 & 24.85 & 17.48 \\
Qwen2-Audio-IT \cite{Qwen2-Audio} & 7B & 67.27 & 56.29 & 33.33 & 24.27 \\
SALMONN \cite{tang2024salmonn} & 13B & 41.14 & 37.13 & 30.30 & 31.07 \\
GAMA-IT \cite{ghosh2024gama} & 7B & 30.93 & 26.74 & 22.42 & 16.02 \\
LTU \cite{gong2024listen} & 7B & 20.42 & 15.97 & 19.39 & 19.90 \\ \hline \hline
CLAP (ours) & 199M & 58.86 & 46.41 & 43.03 & 26.11 \\
CLAP+AudioMCQ & 199M & 64.56 & 52.99 & 45.45 & 31.03 \\
TP-CLAP (proposed) & 219M & \textbf{71.47} & \textbf{55.99} & \textbf{47.88} & \textbf{32.02} \\ \hline
\end{tabular}
}
\vspace{-3mm}
\end{table}

Table~\ref{tab:audiomcq_sources} compares TP-CLAP with representative audio-LLMs on MMAU and MMAR. These audio-LLMs combine pretrained audio encoders and LLMs and are instruction-tuned on diverse audio understanding tasks, making them well suited for question answering. In contrast, TP-CLAP is built upon a lightweight CLAP architecture designed for contrastive representation learning and retrieval.

Due to strong audio--language alignment, the base CLAP model already achieves competitive performance with several audio-LLMs, such as SALMONN~\cite{tang2024salmonn} and GAMA-IT~\cite{ghosh2024gama}. Incorporating AudioMCQ supervision (CLAP+\allowbreak AudioMCQ) further improves performance over the base model. The proposed fusion module then provides additional gains across all evaluation sets. These improvements stem from the cross-attention mechanism, which enables text-conditioned reweighting of audio features. Unlike CLAP+\allowbreak AudioMCQ, which relies on fixed audio embeddings, TP-CLAP dynamically adapts audio representations to the input question, allowing more selective focus on task-relevant acoustic cues.

As expected, Audio Flamingo 3~\cite{ghosh2026audio} achieves the strongest overall performance owing to its substantially larger model capacity and scale of training data. Nevertheless, TP-CLAP performs competitively with other selected audio-LLMs, particularly on the sound subsets, despite using only 219M parameters. This highlights the effectiveness of contrastive representations for audio understanding without relying on an autoregressive language model.

\subsection{Attribute-Focused Audio-to-Audio Retrieval}


\begin{table}[t]
\centering
\caption{Attribute-focused retrieval results on NSynth.}
\label{tab:nsynth_results}
\begin{tabular}{l|l|ccc|c}
\hline
Attribute & Setting & P@5 & P@10 & P@20 & mAP \\
\hline
\multirow{2}{*}{Instrument}
& Uncon. & 57.0 & 46.1 & 38.0 & 22.0 \\
& Con. & \textbf{77.8} & \textbf{77.6} & \textbf{76.3} & \textbf{45.0} \\
\hline
\multirow{2}{*}{Pitch}
& Uncon. & 75.3 & 72.2 & 70.0 & 60.2 \\
& Con. & \textbf{88.0} & \textbf{86.8} & \textbf{84.3} & \textbf{72.2} \\
\hline
\multirow{2}{*}{Source}
& Uncon. & 65.2 & 56.7 & 51.1 & 39.0 \\
& Con. & \textbf{78.7} & \textbf{78.8} & \textbf{78.7} & \textbf{58.2} \\
\hline
\end{tabular}
\vspace{-4mm}
\end{table}

\begin{table}[t]
\centering
\caption{Attribute-focused retrieval results on MagnaTagATune.}
\label{tab:mtt_results}
\begin{tabular}{l|l|ccc|c}
\hline
Attribute & Setting & SP@5 & SP@10 & SP@20 & SmAP \\
\hline
\multirow{2}{*}{Genre}
& Uncon. & 65.4 & 64.8 & 64.1 & 55.2 \\
& Con. & \textbf{66.3} & \textbf{66.4} & \textbf{66.3} & \textbf{58.0} \\
\hline
\multirow{2}{*}{Instrument}
& Uncon. & 66.8 & 66.0 & 65.2 & 50.9 \\
& Con. & \textbf{69.7} & \textbf{70.0} & \textbf{69.5} & \textbf{54.9} \\
\hline
\multirow{2}{*}{Tempo}
& Uncon. & 93.4 & 92.8 & 91.4 & 81.1 \\
& Con. & \textbf{96.5} & \textbf{96.5} & \textbf{96.6} & \textbf{90.2} \\
\hline
\end{tabular}
\vspace{-3mm}
\end{table}

Tables~\ref{tab:nsynth_results} and~\ref{tab:mtt_results} present attribute-focused retrieval results on the NSynth and MagnaTagATune test sets. Retrieval using the unconditioned TP-CLAP embedding is compared with retrieval using the text-conditioned representation obtained from the same model. The text-conditioned representation outperforms the unconditioned one across all evaluated attributes and metrics, showing that TP-CLAP successfully adapts audio representations to the requested attribute.

Importantly, on both datasets, the unprompted retrieval performance consistently decreases from (S)P@5 to (S)P@20, whereas prompted retrieval remains comparatively stable across different retrieval depths. This suggests that the global CLAP embedding captures multiple acoustic factors simultaneously, causing retrieval rankings to become increasingly influenced by attributes unrelated to the target criterion as more items are retrieved. In contrast, text-conditioned representations emphasise the requested attribute, resulting in a more coherent ranking of relevant samples.

\section{CONCLUSION}

This paper proposes TP-CLAP, a lightweight extension of CLAP that enables text-conditioned audio representations through a cross-attention-based fusion module. Using AMCQA supervision, TP-CLAP incorporates textual prompts into audio embeddings while remaining compatible with the original CLAP framework. Experiments demonstrate that TP-CLAP performs competitively with several substantially larger audio-LLMs on audio question answering. It further enables attribute-focused audio-to-audio retrieval, where text-conditioned representations consistently outperform unconditioned ones, allowing retrieval to be guided by user-specified characteristics. In addition, TP-CLAP improves audio–text retrieval and zero-shot classification over the base CLAP model.


\bibliographystyle{IEEEbib}
\bibliography{refs}

\end{document}